# Analysis and Visualization of the Parameter Space of Matrix Factorization-based Recommender Systems

Hao Wang*[a]
[a] Ratidar Technologies LLC, Beijing, China, 100011
* Corresponding author: haow85@live.com

**ABSTRACT**

Recommender system is the most successful commercial technology in the past decade. Technical mammoth such as Temu, TikTok and Amazon utilize the technology to generate enormous revenues each year. Although there have been enough research literature on accuracy enhancement of the technology, explainable AI is still a new idea to the field. In 2022, the author of this paper provides a geometric interpretation of the matrix factorization-based methods and uses geometric approximation to solve the recommendation problem. We continue the research in this direction in this paper, and visualize the inner structure of the parameter space of matrix factorization technologies. We show that the parameters of matrix factorization methods are distributed within a hyper-ball. After further analysis, we prove that the distribution of the parameters is not multivariate normal.

## 1. INTRODUCTION

Before the invention of ChatGPT, recommender system is considered as a potent competitor to search engine. Major internet players such as Temu, TikTok, JD.com and Amazon.com spend lavishly on R&D of the technology. Unlike other technologies such as programming languages (Python / Go / Julia / Clojure) or databases, recommender system could save billions of USD per year for the industry. By hiring only a team of less than 1000 people, large corporations could increase the volume of traffic or sales (Toutiao.com / Amazon.com) by 30% - 40% without spending marketing fees using Google Ads. The business is extremely profitable for internet firms, and even after decades of evolution, recommender systems is still not wiped into the dust bin of history.

One of the most successful recommender system technologies is matrix factorization. Due to its versatile functionality which could incorporate feature engineering and be easily transformed into online learning, matrix factorization is widely adopted in the industry. The most commonly used optimization technique used to solve matrix factorization problem is Stochastic Gradient Descent. With a small randomly sampled dataset, we could solve matrix factorization problem with only several lines of Python code and extremely fast execution speed.

Although scientists have shifted their main focus from shallow models such as matrix factorization into deep neural models such as Wide & Deep [1], we are still far from being safe to claim that we fully understand the mechanism behind the matrix factorization framework. For example, although Probabilistic Matrix Factorization [2] was invented as early as 2007, it was only until 2021 and 2022 that true zeroshot learning algorithms [3][4][5][6] based on matrix factorization were proposed. The time gap is 14 years. In 2022, an explainable AI paper [7] was published, which was one of the first geometric interpretation of matrix factorization methods. The time gap is 15 years. The majority of scientists and engineers have spent so much time on accuracy improvement, that they neglected the theoretical foundation of the recommender system technologies.

In this paper, we provide an innovative analysis of the matrix factorization-based recommender systems. We choose 2 specific examples of modern day matrix factorization approaches, namely KL-Mat [8] and ZeroMat [3], as test benchmark, to analyze the geometry of the parameter space of the paradigm. We demonstrate in our experiments that the distribution of the parameters of the matrix factorization approaches is spherical in the space, but it is not multivariate normal distribution.

## 2. RELATED WORK

Recommender system was invented in late 1980's. Collaborative filtering was introduced as the first batch of the recommender system, although its distributed versions are introduced much later. In 2007, a milestone paper introducing the probabilistic framework of matrix factorization paradigm was published. The paper provides a theoretical foundation using MAP estimation for matrix factorization-based the recommender system technology. Major innovation ideas for matrix factorization framework including SVDFeature [9], SVD++ [10], timeSVD [11], etc. The early inventions focus on increasing the accuracy performance of the algorithm and embodying more user cases. Later inventions focus on cold-start problem, such as ZeroMat [3], DotMat [4], PowerMat [5] and PoissonMat [6], as well as fairness problem, such as Focused Learning [12], MatRec [13], Zipf Matrix Factorization [14], KL-Mat [8], and RankMat [15]. Matrix factorization algorithms could be also used to solve Context-aware Recommendation problem (CARS).

Linear models is the next revolutionary wave in the field. Netease [16] and Baidu [17][18] took advantage of linear and linear hybrid models to solve the recommendation and personalization problems. Unlike collaborative filtering and matrix factorization models, linear models assume linearity of the problem structures, and is much faster than other nonlinear methodologies.

As the rise of deep neural network models in 2012, recommender system has witnessed a revolution initiated by the technology. Influential models such as DeepFM [19], DLRM [20], and Wide & Deep algorithms [1] have become the de facto standards of industrial engineering of the field. Companies such as Google [21] and Netease [22] spend a lot of resources developing the technologies. There are also researchers who start to research on neural architecture search problem of the field [23][24][25]. Deep neural networks help the internet firms increase their accuracy metrics, but at the same time have increased the cost of infrastructure enormously.

Explainable AI is still a new idea in the field. One notable publication in this subfield is Wang [7], which interprets matrix-factorization based recommender system problem as a geometric approximation problem.

## 3. GEOMETRIC ANALYSIS

The official definition of Matrix Factorization is the following formula :

$$L = \sum_{i=1}^{N} \sum_{j=1}^{M} \left( \frac{R_{i,j}}{R_{max}} - \frac{U_i^T \bullet V_j}{\|U_i^T \bullet V_j\|} \right)^2$$

By Zipf Law, we know that $U_i^T \bullet V_j$ are distributed proportional to $U_i^T \bullet V_j$ , or roughly speaking , the number of vector pairs whose dot product value is N is proportional to N. This parameter space is very complicated by a first look, and it is extremely hard to carry out analytical formulas to depict the space. So we visualize the parameter space of 2 specific examples of matrix factorization paradigm to check the distribution of the parameters. The 2 examples that we pick up are KL-Mat, which is a fairness enhancement technique; and ZeroMat, which is a data agnostic zeroshot learning algorithm.

We test our algorithms on the MovieLens 1 Million Dataset [26], which contains 6040 users and 3706 items. We visualize the results by t-SNE [27] in Fig. 1.

Fig.1 illustrates the visualization of distribution of user feature vectors and item feature vectors in KL-Mat by different Stochastic Gradient Descent steps. As can be observed from the figure, both the user feature vectors (Blue) and the item feature vectors (Green) are distributed within a hyber-ball. The radius of the user feature vector hyper-ball is slightly larger than the item feature vector hyper-ball. We use the Henze-Zirkler Test [28] to check whether the parameter space is multivariate normal and the test fails at every Stochastic Gradient Descent Learning step. The result is pretty shocking, because: 1. The geometry of the parameter space is so simple that it is hard for us to believe the result is correct ; 2. Although the result looks simple on the surface, we are yet not ready to understand the mechanism of the distribution, because it is not normally distributed.

Subplots of Fig.1 demonstrate that although the vectors of the parameter spaces are spherically distributed, they do not seem to vary according to different gradient learning steps. The brim of the difference sets of blue and green points seem to be uniformly distributed, and the core of the intersections of the two sets seem to be denser. The implications of the distribution and a much more rigorous analysis will be found in the Discussion Section.

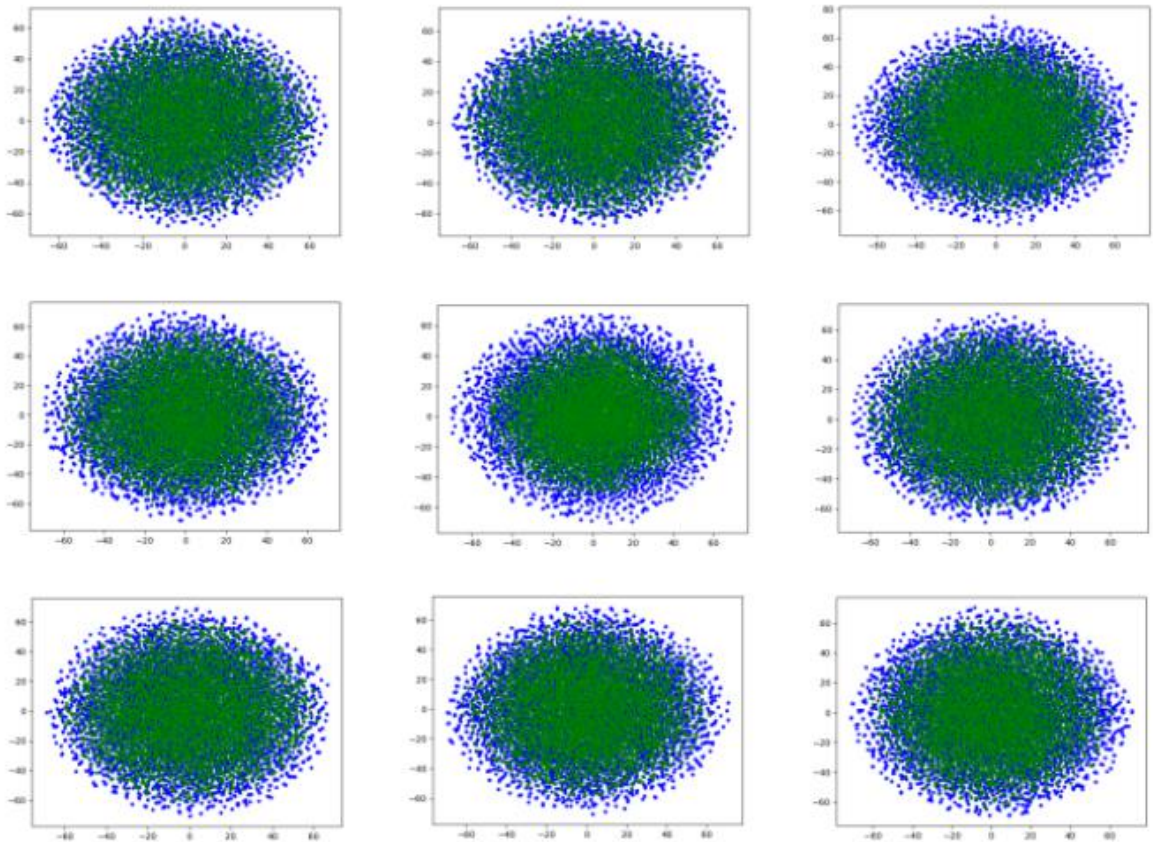

Fig 1. Distribution of User Feature Vectors and Item Feature Vectors in KL-Mat by Different Stochastic Gradient Descent Learning Steps

We now test the zeroshot learning algorithm ZeroMat on the same MovieLens 1 Million Dataset, and obtain the results in Fig. 2.

Fig. 2 illustrates the visualization of distribution of user feature vectors and item feature vectors in ZeroMat by different Stochastic Gradient Descent steps. As can be observed from the figure, both the user feature vectors (Blue) and the item feature vectors (Green) are distributed within a hyber-ball. The radius of the user feature vector hyper-ball is slightly larger than the item feature vector hyper-ball. We use the Henze-Zirkler Test to check whether the parameter space is multivariate normal and the test fails at every Stochastic Gradient Descent Learning step.

The analysis of subplots of Fig.2 is analogous to Fig.1. We leave a formal analysis to the Discussion Section.

## 4. DISCUSSION

We take the specific example of KL-Mat and investigate into the distribution of its user feature vector parameter space. We found that although the 2D histogram of the distribution does not tell us much, by observing the 1D histogram of the X-axis and Y-axis of the t-SNE visualization of the parameter space, we draw the conclusion that the distribution of singular dimensional data of the parameter space is triangular :

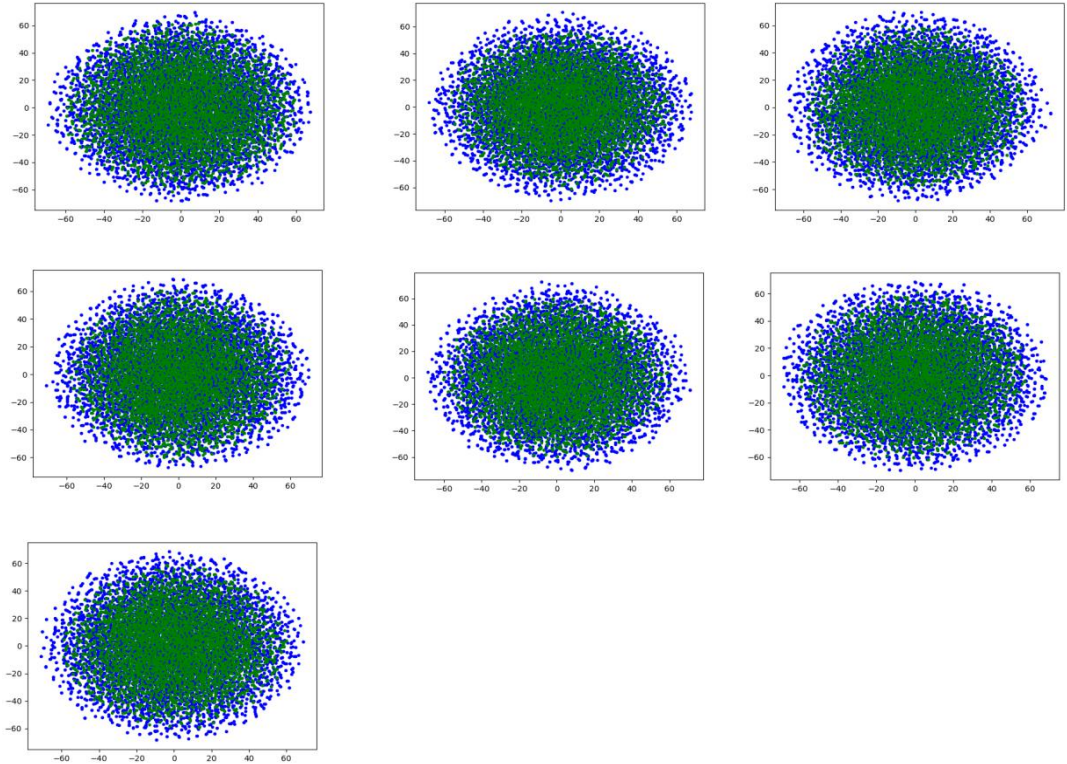

Fig 2. Distribution of User Feature Vectors and Item Feature Vectors in ZeroMat by Different Stochastic Gradient Descent Learning Steps

Fig. 3 demonstrates that the X-axis of the user feature vectors follow triangular distribution. We also checked the Y-axis of the user feature vectors, and the vector feature space, the distributions are uniquely triangular.

Fig. 4 demonstrates that the 3D histogram of the user feature vector density function of KL-Mat is a cone, while the item feature vector density function is similar.

## 5. CONCLUSION

In this paper, we visualized the parameter spaces of KL-Mat and ZeroMat using t-SNE in Fig.1 and Fig.2 and claim that the parameter spaces failed the Henze-Zirkler Multivariate Normality test. After visualizing the histogram of the probability density of KL-Mat and Zero-Mat, we observe that the probability density function of the parameter spaces of matrix factorization methods is a cone. We are surprised the complexity of the parameter spaces leads to a rather simple geometry on appearance, but a rather complicated underlying probabilistic mechanism.

In future work, we would like to find out an analytical form for the probabilistic distribution underlying the spherical distribution and pin down the properties of the matrix factorization parameter space. We would also like to conduct a rigorous hypothesis testing for the triangular property of the probability density function of the parameter spaces of the matrix factorization method.

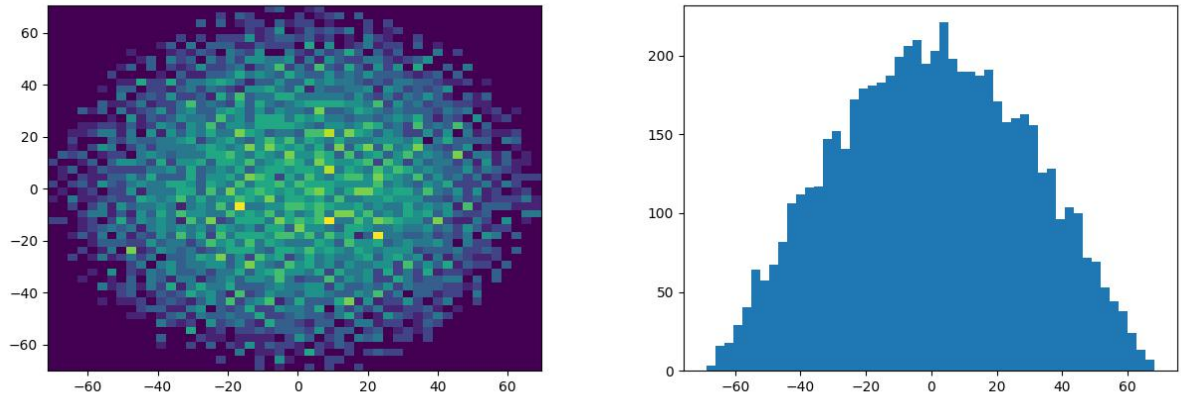

Fig. 3 A Specific Example of KL-Mat : Plot on the Left shows the 2D Histogram ; Plot on the Right shows the 1D Histogram of the X-axis of the t-SNE visualization

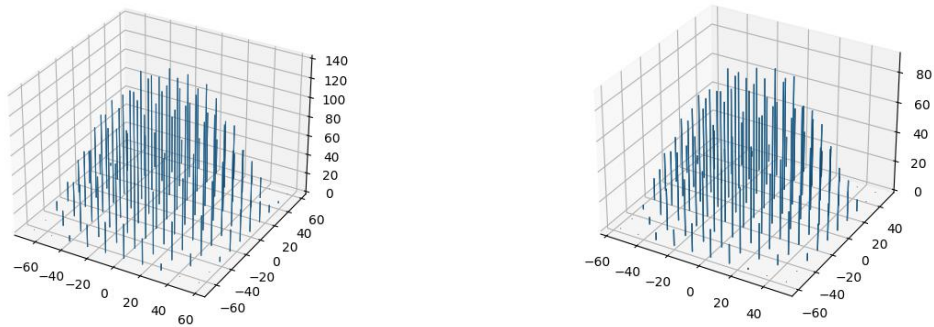

Fig. 4 A Specific Example of KL-Mat : Plot on the Left shows the 3D Histogram of the user feature vectors; Plot on the Right shows the 3D Histogram of the item feature vectors of the t-SNE visualization